\documentclass[aps,prl,reprint,superscriptaddress,showpacs,floatfix]{revtex4-1}
\usepackage{graphicx}
\usepackage{hyperref}
\usepackage{dcolumn}
\usepackage{bm}
\begin{document}
\title{Terahertz Radiation Induced Ballistic Electron Transport in Graphene}

\author{Shuntaro Tani}
\email{tani@scphys.kyoto-u.ac.jp}
\affiliation{Department of Physics, Kyoto University, Sakyo-ku, Kyoto 606-8502, Japan}
\affiliation{CREST, Japan Science and Technology Agency, Kawaguchi, Saitama 332-0012, Japan}

\author{Fran\c{c}ois Blanchard}
\affiliation{CREST, Japan Science and Technology Agency, Kawaguchi, Saitama 332-0012, Japan}
\affiliation{Institute for Integrated Cell-Material Sciences (WPI-iCeMS), Kyoto University, Sakyo-ku, Kyoto 606-8501, Japan}

\author{Koichiro Tanaka}
\email{kochan@scphys.kyoto-u.ac.jp}
\affiliation{CREST, Japan Science and Technology Agency, Kawaguchi, Saitama 332-0012, Japan}
\affiliation{Institute for Integrated Cell-Material Sciences (WPI-iCeMS), Kyoto University, Sakyo-ku, Kyoto 606-8501, Japan}

\date{\today}

\begin{abstract}
We investigated ultrafast carrier dynamics in graphene with near-infrared transient absorption measurement after intense half-cycle terahertz pulse excitation. The terahertz electric field efficiently drives the carriers, inducing large transparency in the near-infrared region. Theoretical calculations using the Boltzmann transport equation quantitatively reproduce the experimental findings. This good agreement suggests that the intense terahertz field should promote remarkable impact ionization process, which leads to suppression of optical phonon emission and results in efficient carrier transport in graphene.
\end{abstract}

\pacs{72.80.Vp, 42.65.-k, 72.20.Ht, 72.40.+w}
\keywords{nonlinear THz response, graphene, ballistic transport}
\maketitle
Graphene possesses many unique properties arising from its truly 2D honeycomb structure, and offers us a solid state playground of Dirac fermions\cite{Novoselov:2005p9768}. One exciting property in graphene is its high mobility even at room temperature owing to its massless band structure and high Fermi velocity\cite{Chen:2008p10401}, leading to long-range ballistic transport \cite{Du:2008p9098}, which is significant for both fundamental physics and industrial applications.
Although great achievements of DC electron transport in graphene \cite{DasSarma:2011p9498,Chen:2008p10401,Du:2008p9098,Berciaud:2010p11192,Meric:2008p9419}, current saturation restricts electron transport under a high electric field with field strength over 10 kV/cm, and has been attributed to optical phonon emission in high-energy state of Dirac electrons\cite{Berciaud:2010p11192,Meric:2008p9419,Barreiro:2009p9415,Tse:2008p9503}. For anticipated graphene based nanoelectronics devices, the phonon scattering may restrict high field operations.

Thanks to recent advances in high-power and time-resolved terahertz (THz) technologies \cite{Hoffmann:2011p11127}, we can apply electric fields of high strength (MV/cm) with a pulse width of $\sim$100 fs to probe the ultrafast nonlinear optical properties of materials in the THz frequency region. For instance, novel carrier dynamics in semiconductors have been reported such as THz-field-driven mass anisotropy in doped InGaAs \cite{Blanchard:2011p11082}, and highly efficient carrier multiplication in GaAs quantum wells \cite{Hirori:2011p9617}, to name only two. In graphene, S.Winnerl et al. has performed THz-pump THz-probe measurement, revealing that the excitation and relaxation dynamics of carriers strongly depend on the excitation frequency in the THz region\cite{Winnerl:2011p10414}. In these studies, THz pulses are of picosecond duration, which is longer than the typical optical phonon emission timescale in graphene. Shorter pulse is required to prevail over current saturation caused by optical phonon emission. 

In this work, we succeeded in ballistic electron transport in graphene using 200-fs half-cycle THz excitation pulses and monitoring their dynamics with 50-fs long near-infrared (NIR) probing pulses. Large THz induced transparency is observed in the NIR region for the first time. Numerical calculations using the Boltzmann equation quantitatively reproduce the experimental findings and demonstrate the importance of efficient carrier multiplication by impact ionization in graphene. This multiplication achieved by ultrashort THz pulse excitation drastically suppresses the current saturation effect.

The sample used in our experiment is a commercial CVD graphene grown on Cu foil, which is transferred onto a $\rm{SiO_{2}}$ substrate. We have confirmed the sample's single-layer nature and uniformity by confocal Raman imaging microscope \cite{Ferrari:2006p9727}. The transmissivity of our sample at 800 nm is 97 $\%$, comparable with the value obtained from universal conductivity of intrinsic single layer graphene \cite{Novoselov:2005p9768}. Because graphene is deposited on $\rm{SiO_2}$, substrate-induced inhomogeneity at the graphene-oxide interface gives rise to p-type doping with Fermi energy over 200 meV \cite{Ryu:2010p10870}.

\begin{figure}[!thb]
\includegraphics{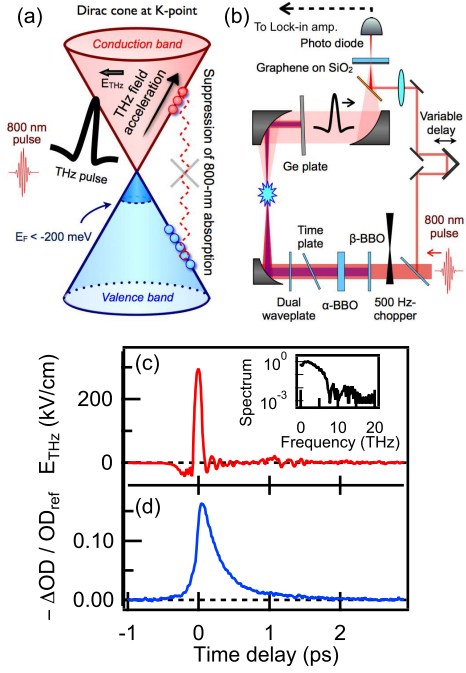}
\caption{(a) Schematic figure of THz-pump NIR-transient absorption measurement in Dirac cone. (b) Experimental setup. (c) Incident THz electric field and its spectrum. (d) THz induced normalized differential optical density of graphene at 800 nm as a function of delay time. }
\label{fig:one}
\end{figure}
To investigate hot electron transport properties in graphene, we performed transient absorption measurement in the NIR region under intense THz excitation, as illustrated in Fig. \ref{fig:one}(a). We used a collinear air plasma method for generation of intense half-cycle THz pulses \cite{Dai:2009p10727} (Fig. \ref{fig:one}(b)). A 800-nm laser pulse (Ti:Sapphire laser that delivers 3 mJ with 35-fs duration laser pulses) was split into two pulses for THz excitation and time-resolved NIR transmissivity measurements. An undoped germanium substrate was inserted in the THz excitation line to reject any remaining 800 nm-pump beam after plasma generation. The laser pump and probe energy are 2.4 mJ and $<$1 nJ per pulse, and spot sizes of THz pulse and 800-nm pulse at the sample position are 80 $\mu$m and 30 $\mu$m, respectively. Figure \ref{fig:one}(c) shows a temporal profile of the THz pulse with a peak electric field of 300 kV/cm and a full-width at half-maximum of $\sim$ 100 fs. Notice that spectrum of THz pulse is ranging from 0.5 to 7 THz (inset of Fig. \ref{fig:one}(c)) with the photon energy several tens of times smaller than that of the probing pulse. All measurements were performed at room temperature and under dehydrated air to suppress THz absorption from water vapor. Temporal evolutions of 800-nm probe transmission through graphene after THz pulse excitation could be reconstructed by scanning an optical delay stage. Finally, a lock-in amplifier (SRS SR830) connected to the output of an amplified photodiode, and referenced to the chopper that switched the excitation on and off at 500 Hz, served to probe the change of the NIR transmissivity $\Delta \rm{T}$.

Figure \ref{fig:one}(d) shows a typical THz induced differential optical density measurement as a function of time with a peak THz electric field of 300 kV/cm. To clarify the physical meaning observed here, we define the normalized differential optical density by $\rm{\Delta OD/OD_{ref} = (OD_{ex} - OD_{ref})/OD_{ref}} \simeq \Delta T/OD_{ref}$ , where $\rm{OD_{ref}}$ is the optical density of graphene at 800 nm without THz pulse excitation, and $\rm{OD_{ex}}$ is that of with THz pulse excitation. The value of $-\rm{\Delta OD/OD_{ref}}$ corresponds to the population occupation at the energy of 800-nm transition. After THz pulse excitation, the differential optical density immediately goes negative  (e.g., graphene becomes transparent in the NIR region under THz pulse excitation) and returns to zero within 2 ps. Experimentally, the maximum value of $\Delta \rm{T}$ reaches $5\times10^{-3}$ for a peak electric field of 300 kV/cm. This large induced transparency over 16$\%$ was probably caused by to electron filling or hole depletion of the corresponding energy levels.

\begin{figure}
\includegraphics[width=8.6cm]{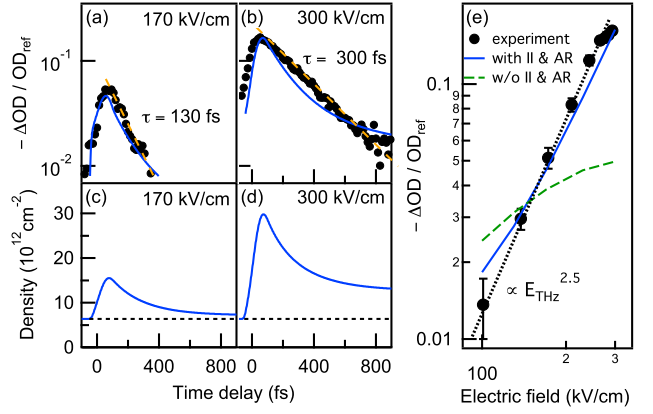}
\caption{THz induced normalized differential optical density as a function of time delay with peak electric fields of (a) 170 kV/cm and (b) 300 kV/cm : experiment (circle) and simulation (solid line). (c)(d) Calculated carrier density as a function of time delay. Dashed lines represent initial carrier density. (e) THz induced transparency of graphene as a function of a peak electric field : experiment (filled circle) and simulations (solid and dashed lines). Experimental results follow a power law with exponent 2.5 (dotted line).}
\label{fig:two}
\end{figure}
Figures \ref{fig:two}(a) and \ref{fig:two}(b) show the temporal profile of induced NIR transparency for peak electric fields of 170 kV/cm and 300 kV/cm. The maximum values of induced transparency drastically increase for the higher electric field, while  carrier relaxation time tends to increase linearly with field strength. The origins leading to the nonlinearity will be discussed in subsequent paragraphs.

Figure \ref{fig:two}(e) shows the maximum values of induced transparency as a function of a peak electric field. A power law with exponent 2.5 is found experimentally. This is in contrast with quadratic dependence observed in ultrafast optical pump-probe experiments \cite{Dawlaty:2008p9699,Carbone:2011p9401}. The super-quadratic dependence found here suggests nonlinear coupling between the THz field and electronic system, and can be explained by the rapid increase in the number of carriers with field strength.

\begin{figure*}[bht!]
\includegraphics[width=15cm]{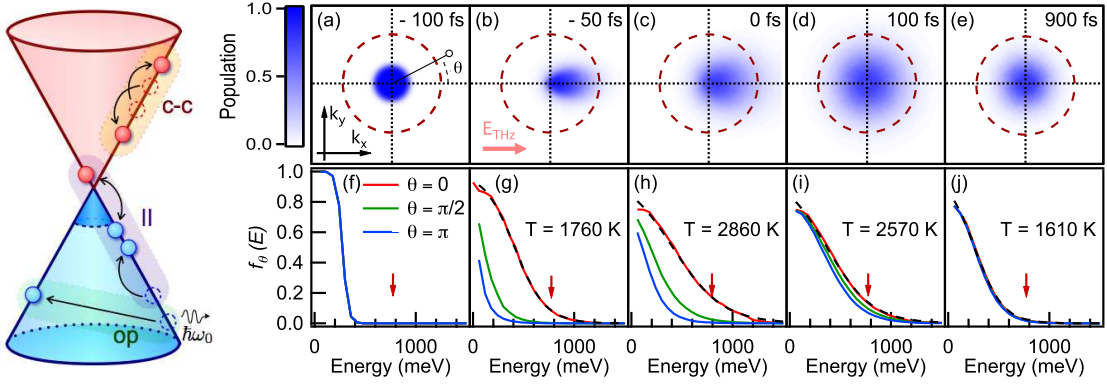}
\caption{(a)-(e) Distribution function of holes in k-space with various time delays. Dashed lines represent momenta corresponding to the energy of 800-nm transition. (f)-(j) Angle-resolved energy distribution of holes. Fermi-Dirac fitting is represented as a dashed line. Brown arrows indicate the energy of 800-nm transition.}
\label{fig:four}
\end{figure*}
As examined in detail in previous studies from THz to NIR regions \cite{Winnerl:2011p10414,Mak:2008p11215,Sun:2008p9700}, electromagnetic wave excitation results in two possible scenarios:  interband and intraband transitions. Interband transition creates electron-hole pairs with kinetic energy of the incident photon energy, whereas intraband transition induces athermal distribution of carriers in the momentum space. According to these studies, the dominant mechanism should be determined by photon energy and Fermi energy of graphene. In this study, the dominant mechanism is restricted to intraband transition because of large negative Fermi energy below -200 meV coming from the $\rm{SiO_2}$ substrate. Since the excited carriers have the kinetic energy corresponding to NIR excitation, their relaxation dynamics can be compared with photo-excitation studies such as ultrafast saturation absorption\cite{Xing:2010p9088}, single-color and two-color pump-probe measurement\cite{Kampfrath:2005p9411,George:2008p8904,Dawlaty:2008p9699,Carbone:2011p9401,Breusing:2011p9705}, and ultrafast photoluminescence measurement\cite{Lui:2010p10412}, where 8 fs carrier-carrier scattering, 50-120 fs carrier thermalization, 0.1-1 ps carrier cooling and 3 - 15 ps carrier recombination have been reported .

To understand the carrier dynamics based on our observations, we have calculated the distribution function of carriers in \textbf{k}-space using the following semi-classical Boltzmann equation:
\begin{eqnarray}
\frac{\partial}{\partial t}f_{\lambda}(\boldsymbol{k}, t) = &-& \frac{q_{\lambda}}{\hbar}\boldsymbol{E}(t)\frac{\partial}{\partial \boldsymbol{k}}f_{\lambda}(\boldsymbol{k}, t)\nonumber\\
&+& \int d\boldsymbol{k}' \bigl\{S_{\boldsymbol{k},\boldsymbol{k}'}^{\lambda-\lambda}+S_{\boldsymbol{k},\boldsymbol{k}'}^{\lambda-ph}-S_{\boldsymbol{k}',\boldsymbol{k}}^{\lambda-\lambda}-S_{\boldsymbol{k}',\boldsymbol{k}}^{\lambda-ph}\bigr\}\nonumber\\
&+&\int d\boldsymbol{k}'\bigl\{
S_{\boldsymbol{k},\boldsymbol{k}'}^{II,\lambda}+S_{\boldsymbol{k},\boldsymbol{k}'}^{AR,\lambda}
-S_{\boldsymbol{k}',\boldsymbol{k}}^{II,\lambda}-S_{\boldsymbol{k'},\boldsymbol{k},}^{AR,\lambda}
 \bigr\}\nonumber,
\end{eqnarray}
where $f_{\lambda}(\boldsymbol{k}, t)$ is a distribution function of electrons and holes ($\lambda = e,h$), $q_{\lambda}$ is the charge of carrier, and $\boldsymbol{E}(t)$ is the incident THz electric field on graphene at time $t$. $S_{\boldsymbol{k},\boldsymbol{k}'}^{\lambda-\lambda}$ represents intraband carrier-carrier (c-c) scattering rates with coulomb interactions \cite{Girdhar:2011p9534}. Intervalley K-K' scattering is neglected for large momentum exchange \cite{Malic:2011p9678}. $S_{\boldsymbol{k},\boldsymbol{k}'}^{\lambda-ph}$ represents intra- and inter-band phonon scattering rate with $\rm{\Gamma}$ and K optical phonon modes \cite{Malic:2011p9678}. Electron-phonon coupling strength is from Ref. \cite{Piscanec:2004p9709}. $S_{\boldsymbol{k},\boldsymbol{k}'}^{II,\lambda}$ and $S_{\boldsymbol{k'},\boldsymbol{k},}^{AR,\lambda}$ represent impact ionization (II) and Auger recombination (AR) rates \cite{Rana:2007p9472}, respectively. The last two terms (II and AR) change the number of carriers through interband c-c scattering. The II process creates an extra electron-hole pair while losing the kinetic energy of another electron or hole, as schematically shown in Fig. \ref{fig:four}. The AR is the inverse process of II, which reduces the number of carriers. The first term represents carrier acceleration on linear dispersion relation $E=\hbar v_F |\boldsymbol{k}|$ (Dirac cone) with an electric field, where $E$ is the energy of the carrier and $v_F$ is the Fermi energy in graphene. In the second term, momentum-resolved hot phonon effects are included \cite{Malic:2011p9678}. Acoustic phonon scattering negligibly contributes in our observation time scale and was not included in our calculations \cite{Winnerl:2011p10414,Malic:2011p9678}.

The occurrence of II and AR in graphene is controversial because of restrictions from energy and momentum conservation laws in the Dirac cone. Although some authors have stated that these processes are prohibited because of a limited number of possible initial and final momentum configurations\cite{Foster:2009p10704}, some experimental and theoretical studies have indicated the possibility of using these processes\cite{Winnerl:2011p10414,Girdhar:2011p9534,Rana:2007p9472,George:2008p8904,Breusing:2011p9705,Winzer:2010p9444}. Here, we allow these processes by loosening the momentum conservation law by $\Delta k = 2\times 10^4$ $\rm{cm^{-1}}$ ($\sim$ 1 meV in kinetic energy). This may come from imperfection of our CVD graphene, which possibly results in larger II and AR processes \cite{George:2008p8904,Grein:2003p10816}.

Calculated induced NIR transparency is plotted as a function of a peak electric field in Fig. \ref{fig:two}(e) (solid line). The experimental results are quantitatively reproduced when II and AR processes are included. For comparison, the calculations by strictly applying conservation laws (without II and AR) are also plotted as a dashed line. In the latter case, the sub-linear scaling with field strength clearly deviates from the experiment for the higher electric field because of rapid optical phonon scattering. We have to stress here that the only adjustable parameter is the initial Fermi energy, which is tuned to reproduce the magnitude of induced transparency in Fig. \ref{fig:two}(e). The best-fitted Fermi energy is -280 meV, which is a reasonable value for graphene on a $\rm{SiO_2}$ substrate\cite{Ryu:2010p10870}. Note that the changes of the initial Fermi energy only offset the field dependence without affecting the power law. This good agreement indicated that efficient carrier multiplication by II prevails over AR under the high electric field condition.

To emphasize the effect of carrier multiplication, we plot the calculated carrier density as a function of time in Figs. \ref{fig:two}(c) and \ref{fig:two}(d). The number of carriers becomes almost five times larger than the initial number of carriers with a peak electric field of 300 kV/cm, resulting in the observed large induced transparency. Recovery of the absorption change at 800 nm is also calculated and shown as solid lines in Figs. \ref{fig:two}(a) and \ref{fig:two}(b). Our calculations qualitatively reproduce the field dependent relaxation time, which confirm the longer relaxation time at higher excitation density. This is due to the hot phonon effect \cite{Sun:2008p9700,Malic:2011p9678}.

Figures \ref{fig:four}(a)-(e) show calculated distribution of holes in \textbf{k}-space at various time delays for the maximum applied electric field of 300 kV/cm. Figures \ref{fig:four}(f)-(j) show the corresponding angle-resolved energy distribution of holes defined as $f_{\theta}(E) = \langle f_h(E,\phi) \rangle_{\theta-\pi/4 < \phi < \theta+\pi/4}$ with the angle $\theta$ in Fig. \ref{fig:four}(a). Before THz pulse excitation, holes are distributed within Fermi energy with 300 K thermal fluctuation (Figs. \ref{fig:four}(a) and \ref{fig:four}(f)). As a THz electric field is applied, the carrier distribution becomes an asymmetric shape (Figs. \ref{fig:four}(b) and \ref{fig:four}(g)), and keeps its asymmetry until the peak electric field (Figs. \ref{fig:four}(c) and \ref{fig:four}(h)), although the asymmetry decreases rapidly by optical phonon scattering. After the THz excitation pulse is passing through the sample, the distribution immediately recovers its symmetric shape (Figs. \ref{fig:four}(d) and \ref{fig:four}(i)) and returns back to the initial distribution by emitting optical phonons (Figs. \ref{fig:four}(e) and \ref{fig:four}(j)). This argument is well confirmed by angle-resolved energy distributions fitted with Fermi-Dirac distribution in the THz field direction $\rm{k_x}$ (Figs. \ref{fig:four}(f)-(j)). For instance, temperature in $\rm{k_x}$ direction is 300 K before excitation and is increased to 2860 K after THz pulse excitation, though temperature in other direction remains low even at the peak electric field (Fig. \ref{fig:four}(h)). After THz pulse excitation, energy distributions immediately thermalize in all directions and return back to the initial distribution through phonon emission. Notice that the dashed circles in Figs. \ref{fig:four}(a)-(e) represent momenta corresponding to the energy of 800-nm transition, also indicated with arrows in Figs. \ref{fig:four}(f)-(j).

The characteristic feature of half-cycle THz pulse excitation is field-induced asymmetry in the momentum distribution, which induces current in the direction of the THz electric field. In the linear dispersion relation, a carrier with the momentum direction $\theta$ induces current by $q_{\lambda}v_F \cos(\theta)$. With this expression, we can estimate how the scattering processes affect current. In short, optical phonon scattering efficiently damps current, whereas c-c scattering, II and AR processes conserve current.

In contrast to high DC electric field studies in which rapid optical emission restricts current flow \cite{Berciaud:2010p11192,Meric:2008p9419,Barreiro:2009p9415,Tse:2008p9503}, current is not saturated under intense THz pulse excitation. This can be understood as follows: In our high field conditions, carriers can gain enough kinetic energy to induce impact ionization before optical phonon emission. Because impact ionization process efficiently reduces kinetic energies of carriers while conserving current, carriers have less chance to emit optical phonons. This suppresses the current saturation and leads to longer ballistic transport of carriers in graphene under high electric fields.

In conclusion, we reported the first observation of large THz-induced near-infrared transparency in graphene. Our results were compared with simulations using the Boltzmann equation and we proved that carrier multiplication occurs in graphene driven by high field and ultrashort THz pulse. The observed dependence of the rise in charge density on THz peak field strength indicates an interband impact ionization mechanism in graphene. The impact ionization process may suppress optical phonon emission and lead to ballistic carrier transport in graphene.

\begin{acknowledgments}
We thank G.Asai for the preparation technique of our sample. This work was supported by Grant-in-Aid for Scientific Research on Innovative Area 'Optical science of dynamically correlated electrons (DYCE)' (Grant No. 20104007), and Grant-in-Aid for Scientific Research (A) (Grant No. 23244065).
\end{acknowledgments}

\end{document}